\documentclass[12pt]{article}
\usepackage{amsfonts,amsmath,amssymb}
\usepackage{scalerel}[2016/12/29]
\usepackage{graphicx}
\usepackage{tikz}

\def\be{\begin{equation}}
\def\ee{\end{equation}}
\def\bea{\begin{eqnarray}}
\def\eea{\end{eqnarray}}

\def\sh{\hskip -0.05cm} 
\def\sb{\hskip -0.1cm} 
 
\def\nn{\nonumber}

\def\rme{\mathrm{e}}
\def\rmi{\mathrm{i}}

\usepackage{amsmath}
\usepackage{epsfig,graphicx}
\usepackage{graphicx}
\usepackage{tikz}

\def\be{\begin{equation}}
\def\ee{\end{equation}}
\def\bea{\begin{eqnarray}}
\def\eea{\end{eqnarray}}

\def\rme{\mathrm{e}}
\def\rmi{\mathrm{i}}

\begin{document}

\thispagestyle{plain}

\title{\bf\Large A shifted binomial theorem and trigonometric series}

\author{St\'ephane Ouvry$^*$ \  {\scaleobj{0.9}{\rm and}} Alexios P. Polychronakos$^\dagger$}

\date{\today}

\maketitle

\begin{abstract}
We introduce a shifted version of the binomial theorem, and use it to study some remarkable trigonometric integrals and their explicit rewriting in terms of binomial multiple sums. Motivated by the expressions  of area generating functions arising
in the counting of closed walks on various lattices, we propose similar sums involving fractional values of the area and show that
they are closely related to their integer counterparts and lead to rational sequences converging to powers of $\pi$. Our results, other than their mathematical interest, could be relevant to generalizations of statistical mechanical models of the
Heisenberg chain type involving higher spins or $SU(N)$ degrees of freedom.
\end{abstract}

\noindent
* LPTMS, CNRS,  Universit\'e Paris-Sud, Universit\'e Paris-Saclay,\\ {$\left. \right. $}\hskip 0.3cm 91405 Orsay Cedex, France; {\it stephane.ouvry@u-psud.fr}

\noindent
$\dagger$ Department of Physics, City College of New York, NY10031 and
the Graduate Center\\ {$\left. \right. $}\hskip 0.3cm of CUNY, New York, NY 10016, USA;
{\it apolychronakos@ccny.cuny.edu} 
\vskip 1cm
\section{Introduction}

\noindent The study of the statistics of  lattice random walks is a subject rich with challenges and rewards.
Closely related to the celebrated Hofstadter model \cite{Hof} and its ``butterfly'' spectrum, it has led to exact expressions for the  algebraic area 
counting of walks involving remarkable trigonometric sums \cite{nous} and to intriguing connections with exclusion statistics
\cite{poly}. The mathematical and physical content of these results seems to be well beyond what has been explored so far.

In \cite{us} we focused on a particular class of objects, namely trigonometric integrals of the type 
\be \int_0^1 dt\prod_{i=1}^j 
 \bigg(2\cos\big(\pi t-\pi(i-1)p/q\big)\bigg)^{r l_i}\label{simple0}\ee where  $l_1,l_2,\ldots,l_j$ is a set of positive or null integers and $r$ a positive integer.  They are  important building blocks of the Hofstadter  model  and
play a key role in the algebraic area enumeration of various types of lattice walks, such as, in particular,
the triangular lattice chiral walks introduced in \cite{poly}. The integers $l_i$ are a generalized composition (called $g$-composition \cite{us}) of an integer $n$ representing the length of the walk, and a summation over their possible values is part of the process, although they have no immediate impact in the analysis of the above sums.

A key ingredient for obtaining the algebraic area combinatorics of such walks was the rewriting of the trigonometric integrals (\ref{simple0}) in terms of the
binomial multiple sums 
 \be \sum_{k_3= -{rl_3/ 2}}^{{rl_3/2} 
 }\ldots \sum_{k_{j}= -{rl_j/2}}^{{rl_{j}/ 2}} {rl_1\choose {rl_1/ 2} +A/2+\sum_{i=3}^{j}(i-2)k_i}{rl_2\choose {rl_2/2} -A/2 - \sum_{i=3}^{j}(i-1)k_i}\prod_{i=3}^{j}{rl_i\choose {rl_i/ 2}+k_i}\label{ouf}\ee
summed with weight  $\rme^{ \rmi \pi A p/q}$ over a variable $A$, which ends up being the algebraic area; namely
\begin{align}
&\int_0^1 dt\prod_{i=1}^j 
 \bigg(2\cos\big(\pi t-\pi(i-1)p/q\big)\bigg)^{r l_i}\nonumber\\
&= \sum_A
 \, \rme^{ \rmi \pi A p/q}\, \nonumber\\&\sum_{k_3= -{rl_3/ 2}}^{{rl_3/2} 
 }\ldots \sum_{k_{j}= -{rl_j/2}}^{{rl_{j}/ 2}} {rl_1\choose {rl_1/ 2} +A/2+\sum_{i=3}^{j}(i-2)k_i}{rl_2\choose {rl_2/2} -A/2 - \sum_{i=3}^{j}(i-1)k_i}\prod_{i=3}^{j}{rl_i\choose {rl_i/ 2}+k_i}\nonumber\\\label{sonice}
\end{align}
The allowed values of the summed 
variable $A$ in (\ref{sonice})  are dictated by the condition that the binomial entries ${rl_1/ 2} +A/2+\sum_{i=3}^{j}(i-2)k_i$ and $ {rl_2/2} -A/2-\sum_{i=3}^{j}(i-1)k_i$ be non-negative integers for all  $k_i\in[-rl_i/2,rl_i/2]$, with $i=3,\ldots,j$. It follows that when  $r$ is even, since then the $k_i$'s are all integers,  $A$  has  to be even;
 when  $r$ is odd, since then the $k_i$'s can be  either integers or half-integers,   $l_1+l_2+\ldots+l_j$ has  to  be even  and $A$ of the same parity as $l_1+l_3+\ldots$ (or equivalently $ l_2+l_4+\ldots$ since $l_1+l_2+\ldots+l_j$ is even). In both cases the summation range  is $A\in[-(g-1)r\lfloor (l_1+\ldots+l_j)^2/4\rfloor, (g-1)r\lfloor (l_1+\ldots+l_j)^2/4\rfloor\;]$.
 
 \noindent Note that  in the limit $q\to\infty$  the  integral becomes trivial
   \begin{align}\int_0^1 dt \big(2\cos(\pi t)\big)^{r(l_1+l_2+\ldots+l_j)}={r(l_1+l_2+\ldots+l_j)\choose r(l_1+l_2+\ldots+l_j)/2}\label{ok} \end{align}
with   an overall binomial counting as output. Note also that the multiple binomial sums in (\ref{ouf})  are  $A\to -A$ symmetric (changing $A$ for $-A$ does not affect it since one can harmlessly replace the $k_i$'s by  $-k_i$) meaning that the $\rme^{ \rmi \pi A p/q}$ expansion in (\ref{sonice}) is
actually a $\cos( \pi A p/q)$ expansion.

From now on we focus on the trigonometric integral (\ref{simple0})  and the identity (\ref{sonice}),
and  for simplicity we consider the $r$ even case.
A question naturally arising is whether the sum in the RHS of   (\ref{sonice}) generalizes to encompass different types of summands and, correspondingly, new types of walks or physical models. As stressed above, the sum has to be over even values of $A$ ($A/2$ integer). An obvious generalization, then,
is to sums over {\it fractional} values $A/2$, potentially of infinite range. The most obvious such generalization involves summing over {\it odd}
values of $A$ ($A/2$ half-integer), and this will be the focus of our work.  In the process, we will encounter  binomial sums of the type (\ref{binex}) but  {shifted}, i.e., where  $k_i$ is now a half-integer with infinite range instead of an integer. This will lead us to formulate a {\it shifted} version of Newton's binomial theorem (\ref{newton}) where the summation index $k$ can be shifted by any real parameter $s$ off the integers. Further, as already alluded to in \cite{us}, this procedure will allow for rational sequences converging to powers of $\pi$, when $s=1/2$, and to other quantities of interest when  $s$ has other rational values.

The generalizations considered in this paper, given the connection of the summed variable $A$ to the algebraic area of lattice walks, would correspond to walks with non-integer
area in plaquette units. The nature of such nonstandard walks could in principle be unearthed by the properties of the corresponding area enumeration formulae. In the sequel, we will focus on the mathematical aspects of such sums, leaving their physics for future work. In the Conclusions section we
speculate about the possible correspondence of these models with interacting spin systems.

\section{Mapping sums over $A$ odd to integrals \label{section1} }

We start by reviewing the transition from the LHS  trigonometric integral of (\ref{sonice})  to its RHS  binomial multiple  sums. In a nutshell, the basic steps are \cite{us}:
 
 \iffalse\noindent -trivially rewrite  (\ref{simple}) as 
 \be
\int_{0}^{1} dt\prod_{i=1}^j 
 \bigg(2\cos\big(\pi t-\pi(i-1)p/q\big)\bigg)^{r l_i}\label{simple0}\ee\fi
 -Use Newton's binomial theorem for an integer  $l$
 \be (1+x)^l=\sum_{k=0}^l {l\choose k}x^k~\quad \text{or} ~\quad (x^{1/2}+x^{-1/2})^l=\sum_{k=-l/2}^{l/2}{l\choose {l/ 2}+k}x^{k}\label{newton}\ee
 where in the last expression $k$ is integer or half-integer depending on $l$ being even or odd,
 to binomial-expand each $i=1, \ldots, j$ cosine in (\ref{simple0})  as
 \be \bigg(2\cos\big(\pi t-\pi(i-1)p/q\big)\bigg)^{rl_i}=\sum_{k_{i}= -{rl_i/2}}^{{rl_{i}/ 2}}{rl_i\choose {rl_i/ 2}+k_i}\rme^{2 \rmi \pi k_it-2 \rmi \pi k_i(i-1)p/q}\label{binex}\ee
 -Note that $\sum_{i=1}^j k_i$ multiplying $2\rmi\pi t$ in the phase is  an integer to perform the $t$ integration, obtaining a Kronecker delta on
 $\sum_{i=1}^j k_i=0$
  
  \noindent  -Define the new summation variable $A=-2\sum_{i=1}^j k_i(i-1)$
  
  \noindent  -Use the two relations arising from the previous two steps, namely $\sum_{i=1}^j k_i=0$ and $A=-2\sum_{i=1}^j k_i(i-1)$, to eliminate
 $k_1$ and $k_2$ as summation variables in favor of $A$.

  \iffalse
\noindent  One has also  reexpressed the binomial multiple  sum in the third line of (\ref{sonice})
\iffalse
\begin{align}
 \sum_{k_3= -{rl_3/ 2}}^{{rl_3/2} 
 }\ldots \sum_{k_{j}= -{rl_j/2}}^{{rl_{j}/ 2}}{rl_1\choose {rl_1/ 2} +A/2+\sum_{i=3}^{j}(i-2)k_i}{rl_2\choose {rl_2/2} -A/2 - \sum_{i=3}^{j}(i-1)k_i}\prod_{i=3}^{j}{rl_i\choose {rl_i/ 2}+k_i}\label{multiple}\end{align}
 \fi
 as an integral by augmenting the  simple integral in the LHS of (\ref{sonice})  to the double integral \[{1\over 2}\int_0^1
 dt\int_0^2
 dt'\prod_{i=1}^j 
 \Big(2\sin(\pi t+\pi(i-1)t'\big)\Big)^{r l_i}\delta(p/q-t')\]  and using $2\sum_{n=-\infty}^{\infty}\delta(p/q-t'-2n)=\sum_{A=-\infty}^{\infty}\rme^{\rmi \pi A(p/q-t')}$ to obtain
 \begin{align}
 & \sum_{k_3= -{rl_3/ 2}}^{{rl_3/2}}\ldots 
 \sum_{k_{j}= -{rl_j/2}}^{{rl_{j}/ 2}}\sb{rl_1\choose {rl_1/ 2} +\sh A/2+\sum_{i=3}^{j}(i-2)k_i}\sh{rl_2\choose {rl_2/2} -\sb A/2-\sh\sum_{i=3}^{j}(i-1)k_i}\sh\prod_{i=3}^{j}{rl_i\choose {rl_i/ 2}+k_i}\nonumber \\&={1\over 2}\int_0^{2}{dt'}\int_0^{1}{dt}\prod_{i=1}^j \bigg(2\sin\big(\pi t+\pi(i-1)t'\big)\bigg)^{r l_i}e^{i\pi At'} \label{top}\end{align}

\noindent Note that  in the limit $q\to\infty$  
   \begin{align}\int_0^1 dt \big(2\sin(\pi t)\big)^{r(l_1+l_2+\ldots+l_j)}={r(l_1+l_2+\ldots+l_j)\choose r(l_1+l_2+\ldots+l_j)/2}\label{ok} \end{align}
   an overall binomial counting is obtained.
   \fi
\noindent

{In order to be able to perform  the $A$ odd summation instead of the required  $A$ even summation, it is  useful to reverse the above steps and move from the RHS of (\ref{sonice}) to its LHS. This will, then, allow us to express the $A$ odd summation as a generalization of the trigonometric integral (\ref{simple0}).}

Starting from the  $A$ even summation in the RHS of (\ref{sonice}), we re-introduce two additional summation variables $k_1,k_2$
and two Kronecked deltas
\be
%\sum_{k_1,k_2 {\text{ integers}}} 
\sum_{k_1,k_2 \;{\it integers}}
\delta\left(A+2\sum_{i=1}^j (i-1)k_i \right) \delta\left(\sum_{i=1}^j k_j \right)
\label{backin}\ee
Since $k_1,k_2$ do not appear in the summand, summing over them simply enforces the Kronecker deltas and does not change the expression. Summing over $A$ first, however, enforces the first Kronecker delta yielding $A= -2\sum_{i=1}^j (i-1) k_i$, and substituting this in the sum gives
%\[
%\sum_{k_1= -{rl_1/ 2}}^{{rl_1/2} 
% }\ldots \sum_{k_{j}= -{rl_j/2}}^{{rl_{j}/ 2}} \delta\left(\sum_{i=1}^j k_i \right)  \prod_{i=i}^{j}{rl_i\choose {rl_i/ 2}+k_i} \rme^{ -2 \rmi \pi\, i k_i  p/q}
% \]
 \[
\sum_{k_1,k_2  \;{\it integers}}\sum_{k_3= -{rl_3/2}}^{{rl_3/ 2}} \ldots \sum_{k_{j}= -{rl_j/2}}^{{rl_{j}/ 2}} \delta\left(\sum_{i=1}^j k_i \right)  \prod_{i=i}^{j}{rl_i\choose {rl_i/ 2}+k_i} \rme^{ -2 \rmi \pi\, i k_i  p/q}
 \]
Finally, implementing the Kronecker delta through an integration 
\[
 \delta\left(\sum_{i=1}^j k_i \right) = \int_0^1 dt \, \rme^{2 \rmi \pi  t \sum_{i=1}^j k_i}
 =   \int_0^1 dt \prod_{i=i}^j \rme^{2 \rmi \pi  t k_i}
 \]
leads to a sum over binomial terms, one for each $k_i$, of the form
\be
\int_0^1 dt \prod_{i=1}^{j} \sum_{k_i} {rl_i\choose {rl_i/ 2}+k_i}  \rme^{2 \rmi \pi  k_i (t - (i-1) p/q)}
\label{binn}\ee
(a change of variable  $t \to t+ p/q$ was also performed to bring the exponentials to the form involving $(i-1)k_i$ rather than $i k_i$).

In the above expression we deliberately left the summations over $k_i$ unspecified. In fact, all $k_1 , \dots , k_j$ will
take integer values: $k_3, \dots ,k_j$ by assumption, and $k_1 , k_2$ {by necessity as explicitely indicated in (\ref{backin}): $A$ is even so $k_2$ has to be  integer because of the first Kronecker  delta; then $k_1$ also has to be integer because of the second Kronecker delta}. This also fixes their
range to $[-r l_i/2 , r l_i/2 ]$, in which the binomials do not vanish. An application of the binomial theorem (\ref{newton}) for each term involving $k_i$, then, recovers the LHS of (\ref{sonice}). So by summing the binomial multiple sum (\ref{ouf}) over $A$ even with weight $\rme^{ \rmi \pi A p/q} $ we obtained the trigonometric integral (\ref{simple0}), that is, we recovered the identity (\ref{sonice}).

Let us now  perform the same summation of the same binomial multiple  sum (\ref{ouf})
\iffalse
\begin{align} \sum_{k_3= -{rl_3/ 2}}^{{rl_3/2} 
 }\hskip -0.1cm\cdots \hskip -0.1cm\sum_{k_{j}= -{rl_j/2}}^{{rl_{j}/ 2}}{rl_1\choose {rl_1/ 2} +A/2+\sum_{i=3}^{j}(i-2)k_i}{rl_2\choose {rl_2/2} -A/2-\sum_{i=3}^{j}(i-1)k_i}\prod_{i=3}^{j}{rl_i\choose {rl_i/ 2}+k_i}\nonumber \end{align}
 \fi
with the same weight $\rme^{ \rmi \pi A p/q}$ but  over   $A$ {\it odd}  instead of $A$ even. We note that, contrary to the $A$ even case, where the summation range is obviously finite, in the $A$ odd case the summation range is by construction infinite. Indeed, since $A$ is odd, both binomial entries ${rl_1/ 2} +A/2+\sum_{i=3}^{j}(i-2)k_i$ and $ {rl_2/2} -A/2-\sum_{i=3}^{j}(i-1)k_i$   are now half-integers. 

{{We can proceed in the same way as in the $A$ even case: we introduce two Kronecker deltas and two 
summation variables $k_1,k_2$ that now need to be {\it half}-integers to satisfy the Kronecker deltas with odd $A$,
\[
%\sum_{k_1,k_2 {\text{ integers}}} 
\sum_{k_1,k_2=-\infty\atop \;{\it half{\text -}integers}}^\infty
\delta\left(A+2\sum_{i=1}^j (i-1)k_i \right) \delta\left(\sum_{i=1}^j k_j \right)
\]
which again does not change the overall sum. Performing the $A$-summation first and proceeding as before, we end up
with almost an identical expression as in (\ref{binn}) with the only difference that the summations $k_1,k_2$ are
over {\it half}-integer values. The Newton binomial theorem can then be applied only to variables $k_3 , \dots , k_j$
and we end up with the relation
\begin{align}
& \sum_{A=-\infty\atop  A\;{\it odd}}^{\infty}
 \, \rme^{ \rmi \pi A p/q}\nonumber\\& \sum_{k_3= -{rl_3/ 2}}^{{rl_3/2} 
 }\ldots \sum_{k_{j}= -{rl_j/2}}^{{rl_{j}/ 2}}
 {rl_1\choose {rl_1/ 2} +A/2+\sum_{i=3}^{j}(i-2)k_i}{rl_2\choose {rl_2/2} -A/2 - \sum_{i=3}^{j}(i-1)k_i}\prod_{i=3}^{j}{rl_i\choose {rl_i/ 2}+k_i}\nonumber\\
 &=\int_0^1 dt  \sum_{k_1,k_2=-\infty\atop  \;{\it half{\text -}integers}}^{\infty} {rl_1\choose {rl_1/ 2}+k_1}{rl_2\choose {rl_2/ 2}+k_2}\rme^{ 2\rmi\pi k_1 t}\rme^{2\rmi\pi k_2(t-p/q)}\prod_{i=3}^{j}\bigg(2\cos\big(\pi t -\pi(i-1)p/q\big)\bigg)^{rl_i}\label{sonicebis}
\end{align}
which is again $A\to -A$ symmetric, meaning that the $\rme^{ \rmi \pi A p/q}$ expansion is a $\cos( \pi A p/q)$ expansion.

\section{ A  {shifted} binomial theorem \label{section3}}

\noindent In  (\ref{sonicebis}), by insisting on summing the  binomial multiple sum  (\ref{ouf}) weighted by $ \rme^{ \rmi \pi A p/q}$ over $A$ odd instead of $A$ even as in (\ref{sonice}),
{we ended up trading in the trigonometric integral (\ref{simple0})   the first two cosines for the  {\it shifted} binomial  sums
\be
\sum_{k_1=-\infty\atop k_1\;{\it half}{\text -}{\it integer}}^{\infty}{rl_1\choose {rl_1/ 2}+k_1}\rme^{ 2\rmi \pi k_1 t}~, \quad \sum_{k_2=-\infty\atop k_2\;{\it half}{\text -}{\it integer}}^{\infty}{rl_2\choose {rl_2/ 2}+k_2}\rme^{ 2\rmi \pi k_2( t-p/q)}\label{half}\ee
where   $k_1$ and $k_2$ are {shifted} by $1/2$ from their usual integer values  in (\ref{binex}).  

It is natural to  wonder if some  generalization of the Newton binomial theorem (\ref{newton}) or (\ref{binex})
%\be (x^{1/2}+x^{-1/2})^{\;rl_1}=\sum_{k_1=-rl_1/2}^{rl_1/2}{rl_1\choose {rl_1/ 2}+k_1}x^{k_1}\label{halfbis}\ee
 could  hold  when, instead of a finite sum over  integers,  one has  an infinite sum over  half-integers as in (\ref{half}).

Indeed, such a generalization exists and plays a central role in rewriting the sums appearing in this paper.
Specifically, what we shall call the {\it shifted binomial theorem} states that
\iffalse\be
\left(1+e^{\rmi \varphi} \right)^l  = \sum_{k=-\infty}^\infty {l \choose k+s} \rme^{\rmi (k+s) \varphi}~,~~
l~ {\text {integer}} ,~ \varphi, s ~{\text {real}} ,~ -\pi\le \varphi \le \pi 
\label{shifty}\ee
\fi for an integer $l$ and any real number $s$
\be
\left(1+ x\right)^l  = \sum_{k=-\infty}^\infty {l \choose k+s} x^{k+s},~ x=e^{\rmi \varphi},~
 -\pi\le \varphi \le \pi 
\label{shifty}\ee
where  the fractional binomial coefficients
are expressed in terms of $\Gamma$-functions
\be
 {l \choose k+s} = {l! \over \Gamma(k+s+1) \Gamma(l-k-s+1)}
\nonumber \ee
{It  is a deformation by a shift $s$ of the standard Newton theorem (\ref{newton}) with the caveat that $x$ is now restricted to be a phase $e^{\rmi \varphi}$ with $-\pi\le \varphi \le \pi$}.
Indeed we note that the left hand side of (\ref{shifty})  depends only on $\rme^{\rmi \varphi}$, while the right hand side involves
the fractional power $(k+s) \varphi$, for which the range of $\varphi$ is relevant.

{We have not found any {{statement or}} proof of this theorem in the mathematics literature, so we provide below our own proof. To this end,} define the function
\be
S(\varphi) = \rme^{-\rmi s \varphi} ~~{\text{for}}~ -\pi\le \varphi \le \pi\, ,~ ~S(\varphi+2\pi) = S(\varphi)
\nonumber\ee
$S(\varphi)$ is by definition a $2\pi$-periodic function (discontinuous at $\varphi =(2 n+1) \pi$) with discrete Fourier modes for integer $k$
\be
S_k = {1\over 2\pi} \int_{-\pi}^\pi S(\varphi) \rme^{-\rmi k \varphi} d\varphi = {\sin \big(\pi(k+s)\big) \over  \pi(k+s)} ~,~~ 
S(\varphi) = \sum_{k=-\infty}^\infty S_k \, \rme^{\rmi k \varphi}
\nonumber\ee
\iffalse
Then, for $-\pi <\varphi < \pi$
\bea
S(\varphi) \left(1+e^{\rmi \varphi} \right)^l &=& \sum_{n=0}^l {l \choose n} \sum_{k=-\infty}^\infty {\sin \pi(k+s) \over  \pi(k+s)} 
\, \rme^{\rmi (k+n) \varphi}\nn\\
({\text {shift}} ~k \to k-n)~~ &=& {\sin \pi s \over \pi} \sum_{k=-\infty}^\infty (-1)^k\rme^{\rmi k \varphi}  \sum_{n=0}^l {l \choose n} {(-1)^n \over k+s-n}\nn\\
&=& - {\sin \pi s \over \pi} \sum_{k=-\infty}^\infty (-1)^k\rme^{\rmi k \varphi} {l!~ \Gamma (-k-s) \over \Gamma(l-k-s+1)}
\nonumber\eea
where the finite sum over $n$ was explicitly performed. Finally, using
\be 
\Gamma (-k-s) \Gamma(k+s+1) = (-1)^{k-1} \Gamma(s) \Gamma(1-s) = (-1)^{k-1} {\pi \over \sin\pi s}
\nonumber\ee

\fi 

\noindent Then, for $-\pi <\varphi < \pi$
\bea
S(\varphi) \left(1+e^{\rmi \varphi} \right)^l &=& \sum_{n=0}^l {l \choose n} \sum_{k=-\infty}^\infty {\sin \pi(k+s) \over  \pi(k+s)} 
\, \rme^{\rmi (k+n) \varphi}\nn\\
({\text {shift}} ~k \to k-n)~~ &=& {\sin( \pi s) \over \pi} \sum_{k=-\infty}^\infty (-1)^k\rme^{\rmi k \varphi}  \sum_{n=0}^l {l \choose n} {(-1)^n \over k+s-n}\nn\\
&=&  {\sin (\pi s) \over \pi} \sum_{k=-\infty}^\infty (-1)^k\rme^{\rmi k \varphi}(-1)^l {l!~ \Gamma (k+s-l) \over \Gamma(k+s+1)}
\nonumber\eea
where the finite sum over $n$ was explicitly performed. Finally, using
\be 
\Gamma (1-z) \Gamma(z) = {\pi \over \sin(\pi z)}
\nonumber\ee
for $z=k+s-l$
we obtain
\be 
S(\varphi) \left(1+\rme^{\rmi \varphi} \right)^l =  \sum_{k=-\infty}^\infty \rme^{\rmi k \varphi} {l! \over \Gamma (k+s+1) \Gamma(l-k-s+1)}
\nonumber\ee
and since $S(\varphi) = \rme^{-\rmi s\varphi}$ when $-\pi\le \varphi \le \pi$, we recover the shifted binomial theorem (\ref{shifty}) which holds for
general fractional $s$.
 In particular,
a generalized Chu-Vandermonde identity\footnote{In the  $s=1/2$ case it becomes 
\begin{align} &{l_1+l_2\choose l'_1+l'_2}=\sum_{k=-\infty \atop k~{\it {half{\text -}integer}}}^{\infty}{{l_1}\choose {l'_1}+k}{ {l_2} \choose {l'_2}-k }\label{chu}\end{align}}
automatically follows  as
\begin{align} &{l_1+l_2\choose l'_1+l'_2}=\sum_{k=-\infty}^{\infty}{{l_1}\choose {l'_1}+k+s}{ {l_2} \choose {l'_2}-k-s }\nonumber\end{align}

The shifted binomial theorem (\ref{shifty})
can be equivalently restated as
\be
\left(2\cos({\varphi/ 2})\right)^l ={\sum _{k=-\infty\atop k{\it {\;integer\;or\;half{\text -}integer}}}^{\infty}}{l \choose {l/ 2} + k + s} \, \rme^{\rmi(k+s) \varphi} ~,~~ -\pi\le \varphi \le \pi
\label{shiftybis}\ee
where $k$ is integer or half-integer depending on $l$ being even or odd so that $l/2+k$ is always an integer. {Eq. (\ref{shiftybis})  is the shifted version of the RHS of (\ref{newton}) with $x$  replaced by $\rme^{\rmi \varphi}$. Note  that here and in what follows  we  take   $k$  to be integer or half-integer  such that (\ref{shiftybis})  reduces to (\ref{newton}) when $s=0$. However as far as the validity of (\ref{shiftybis}) per se is concerned, this  qualification on $k$ is not needed, i.e., it can   be taken  integer regardless of the parity of $l$.}

The shift $s=1/2$ is the case of interest in view of the {shifted} binomial sums in (\ref{half}). In this case (\ref{shiftybis}) can be subsumed as
(absorbing the $1/2$ in the summation index $k$  and replacing $\varphi \to 2\pi t$)
 \be \sum_{k=-\infty\atop k\;{\it\;{half{\text -}integer\; or\; integer}}}^{\infty}{l\choose {l/ 2}+k}\rme^{2\rmi\pi k t}=\big(2\cos(\pi t )\big)^{\;l}~,~~ -1/2 \le t \le 1/2
 \label{shiftcos}\ee
 with the values of $k$ chosen such that $l/2 + k$ be half-integer. {Both sums in (\ref{half}) are precisely of this type for $l$
 even,  with in the second sum $t\to t-p/q$ with an integration range 
$-1/2 \le t-p/q \le 1/2$  for  (\ref{shiftcos}) to be  valid.}

We will explore the properties of these sums in the upcoming sections. But for now, we can take advantage of the shifted binomial expansion (\ref{shiftcos}) to retrieve rational sequences that converge to powers of $\pi$ (see e.g., \cite{pi} for a catalogue of $\pi$ formulae). 

As a first example,  let us  fix  $t=0$  in (\ref{shiftcos})  for $l$ even:
this would yield
\be 
\sum_{k=-\infty\atop k\;{\it half{\text -}integer}}^{\infty}{l\choose {l/ 2}+k}=2^{\;l} \nonumber\ee
which, when rewritten as
\be 
2^{-l}\, \lim_{m\to \infty}\sum_{k=-m+1/2\atop k\;{\it half}{\text -}{\it integer}}^{m+1/2}\pi{l\choose {l/ 2}+k}\iffalse= 
2^{1-l}\,  \lim_{m\to \infty}\sum_{k=1/2\atop k\;{\it half}{\text -}{\it integer}}^{m+1/2}\pi{l\choose {l/ 2}+k}\fi=\pi
\nonumber \ee
allows for sequences of rational numbers\footnote{The $l$ odd case would give \be 2^{-l}\lim_{m\to \infty}\sum_{k=-m}^{m}\pi{l\choose {l/ 2}+k}\iffalse =2^{1-l}\lim_{m\to \infty}\sum_{k=1}^{m}\pi{l\choose {l/ 2}+k}+2^{-l}\pi{l\choose {l/ 2}}\fi=\pi\nonumber \ee which    happens to be identical to the $l-1$ even sequence  above  since
 \be{l\choose l/2+k}={l-1\choose l/2+k}+{l-1\choose l/2-k}\nonumber\ee}  converging  to $\pi$ when $m\to\infty$, since for $l$ integer and $l'$ half-integer   
\be \pi {l\choose l' }=(-1)^{l'+1/2}\;{l!}\prod_{k=l'}^{l+l'} {1\over l-k }\nonumber\ee
  is  a rational number.

As a second example, we can integrate  both sides of (\ref{shiftcos}) over $t$ in the interval of validity $[-1/2,1/2]$ of the shifted binomial theorem to obtain,
when $l$ is even\footnote{The $l$ odd case   would trivially yield ${l\choose l/2}={l\choose l/2}$  since only the $k=0$ term in the $k$ even summation would contribute.} 
\be \sum_{k=-\infty\atop k\;{\it half}{\text -}{\it integer}}^{\infty}{l\choose {l/ 2}+k}{\sin(\pi k)\over \pi k}=\int_{-1/2}^{1/2}\big(2\cos(\pi t )\big)^{\;l}\label{ceca}\ee 
i.e.,
\be \sum_{k=-\infty\atop k\;{\it half}{\text -}{\it integer}}^{\infty}{l\choose {l/ 2}+k}{(-1)^{k-1/2}\over \pi k}={l\choose l/2}\nonumber\ee 
This, when rewritten  as 
 \be 
{ (l/2)!^2 \over l!}  \lim_{m\to \infty}\sum_{k=-m+1/2\atop k\;{\it half}{\text -}{\it integer}}^{m+1/2}\;\pi{l\choose {l/ 2}+k}{(-1)^{k-1/2}\over k} = \pi^2
 \nonumber \ee
allows for a sequence of rational numbers  converging now to $\pi^2$ when $m\to\infty$.

This construction can be generalized by considering an arbitrary rational $s$ in (\ref{shiftybis})  to obtain rational sequences converging for example to $\pi/\sin(\pi s)$ or  $\pi^2/\sin(\pi s)^2$ instead of $\pi$ or $\pi^2$  sequences just obtained  for $s=1/2$ (see the Appendix for details).

 \section{Exploring (\ref{sonicebis})\label{different}}
  
Since the trigonometric integral (\ref{simple0})  is  clearly unaffected by changing the integration range  from $[0,1]$ to  any interval of length $1$, in particular to the interval $[-1/2,1/2]$, it can be viewed as a generalization of the integral in (\ref{ceca}) to a product of 
$i=1, \ldots, j$ cosines  with  $\pi(i-1)p/q$ phase shifts.
\iffalse, so  that  we might   as well
 end up considering
\be
\int_{-{1\over 2}}^{1\over 2} dt\prod_{i=1}^j 
 \bigg(2\cos\big(\pi t-\pi(i-1)p/q\big)\bigg)^{r l_i}\label{simplebis}\ee
\fi
It is therefore tempting to expect that  by following the same line of reasoning as in  section {\bf 2}, i.e.,  trading in (\ref{simple0})  some  cosines for their {\it shifted} binomial sums, we might be able to construct more general sequences   for other powers of $\pi$.

There are, however, two caveats to this: Firstly, the freedom to change the integration range of the  variable $t$ exists
only when the sum of the shifts in all binomials is an integer. This is the only case in which the sum of all $k_i$ will be
an integer and the integration will simply pick the term where this sum is zero, all other terms vanishing. In the more general case, {\it all} terms will contribute in a way depending on the range of integration.

Secondly, the {\it shifted} binomial theorem  only holds  when the  argument $\varphi$ is confined in the interval $[-\pi,\pi]$,
while an obvious generalization applies for $\varphi$ outside this range. Therefore
it can hold for at most {\it one} of the
cosines traded for a shifted binomial sum, but not  for the remaining ones, because of the varying $-\pi(i-1)p/q$ phase shifts. E.g.,
it will hold for the first $i=1$ traded cosine, provided that the integration range is $[-1/2,1/2]$ such that $2 \pi t$ be
in the interval  $[-\pi,\pi]$, but not  for the $i=2, \ldots, j$ cosines in which their $-\pi(i-1)p/q$ phase shifts will drive their
arguments outside the range $[-\pi,\pi]$.

There are some notable cases where these  caveats are not relevant: if an {\it even} number of cosines are traded with shifts
$1/2$,  as in (\ref{sonicebis}), then, as already stressed above,  the integration range can freely be changed. (When an  odd number of cosines are
traded the results do depend on the range, as in  section {\bf \ref{canne}} below). Further, in the  $q\to\infty$  limit the
$-\pi(i-1)p/q$ phase shifts vanish and the {\it shifted} binomial theorem  holds for all traded cosines, thus recovering (\ref{simple0}) in the $q\to\infty$ limit, i.e., (\ref{ok}).

{Let us explore the properties of (\ref{sonicebis}) in the light of what has just been said. As a first step, we can 
harmlessly change the integration range from $[0,1]$ to $[-1/2,1/2]$ and then use the {\it shifted} binomial theorem to write
 the first  {\it shifted} binomial sum as a cosine. The integral in the RHS of (\ref{sonicebis}) can then be rewritten as 
\begin{align}
&\int_0^1 dt\sum_{k_1=-\infty\atop k_1\;{\it half}{\text -}{\it integer}}^{\infty}\sum_{k_2=-\infty\atop k_2\;{\it half}{\text -}{\it integer}}^{\infty}{rl_1\choose {rl_1/ 2}+k_1}{rl_2\choose {rl_2/ 2}+k_2}\rme^{ 2\rmi\pi k_1 t}\rme^{2\rmi\pi k_2(t-p/q)}\prod_{i=3}^{j}\bigg(2\cos\big(\pi t -\pi(i-1)p/q\big)\bigg)^{rl_i}\nonumber\\
&=
\int_{-{1\over 2}}^{1\over 2} dt\bigg(2\cos\big(\pi t \big)\bigg)^{rl_1}\sum_{k_2=-\infty\atop k_2\;{\it half}{\text -}{\it integer}}^{\infty}{rl_2\choose {rl_2/ 2}+k_2}\rme^{2\rmi\pi k_2(t-p/q)}\prod_{i=3}^{j}\bigg(2\cos\big(\pi t -\pi(i-1)p/q\big)\bigg)^{rl_i}\nonumber
\end{align}
{  Then, we can  proceed as before: binomial-expand as in (\ref{binex}) the first cosine and each of the  $i=3, \ldots, j$ cosines, see that the overall $\sum_{i=1}^j k_i=k_1+k_2+\sum_{i=3}^jk_i$ multiplying $2\rmi\pi t$ in the exponential is  a half-integer,
define $A=-2\sum_{i=1}^j k_i(i-1)=-2k_2-2\sum_{i=3}^j k_i(i-1)$, with $A$ being necessarily odd since $2k_2$ is odd, 
solve for $k_2=-A/2-\sum_{i=3}^{j}(i-1)k_i$ now expressed in terms of $A$ and the $k_i$'s, $i\ne 2$, and finally integrate over $t$ to obtain
\begin{align}&\int_{-{1\over 2}}^{1\over 2} dt\bigg(2\cos\big(\pi t\big)\bigg)^{rl_1}\sum_{k_2=-\infty\atop k_2\;{\it half}{\text -}{\it integer}}^{\infty}{rl_2\choose {rl_2/ 2}+k_2}\rme^{ 2\rmi \pi k_2 (t-p/q)}\prod_{i=3}^{j}\bigg(2\cos\big(\pi t -\pi(i-1)p/q\big)\bigg)^{rl_i}\nonumber\\=
&\sum_{A\;\text{odd}}
 \, \rme^{ \rmi \pi A p/q}\sum_{k_1=-{rl_1/ 2} }^{{rl_1/2} 
 }\sum_{k_3= -{rl_3/ 2} }^{{rl_3/2} 
 }\ldots \sum_{k_{j}= -{rl_j/2} }^{{rl_{j}/ 2}} \,{\sin\big(\pi(A/2+\sum_{i=1}^{j}(i-2)k_i)\big)\over \pi(A/2+\sum_{i=1}^{j}(i-2)k_i)} \nonumber\\&{rl_1\choose {rl_1/ 2}+k_1} {rl_2\choose {rl_2/ 2} -A/2-\sum_{i=3}^{j}(i-1)k_i}\prod_{i=3}^{j}{rl_i\choose {rl_i/ 2}+k_i}\label{catiede}
\end{align}
which is again $A\to-A$ symmetric 
because the ratio ${\sin\big(\pi(A/2+\sum_{i=1}^{j}(i-2)k_i)\big)\over \pi(A/2+\sum_{i=1}^{j}(i-2)k_i)}$ remains unchanged by simutaneously trading $A$ for $-A$ and the $k_i$'s for $-k_i$. It follows  that the $\rme^{ \rmi \pi A p/q}$ expansion (\ref{catiede}) is in fact a $\cos(\pi A p/q)$ expansion.
Both expansions in (\ref{sonicebis}) and  (\ref{catiede})  are identical,
meaning that for a given $A$ their respective   multiple binomial sums weighted by $\rme^{ \rmi \pi A p/q}$, which in both cases have finite summation range and are of the form $1/\pi^2\;\;\times$ a rational number\footnote{In (\ref{sonicebis}) the first two binomials have half-integer entries, and in (\ref{catiede}) only the second binomial has a half-integer entry but there is an additional factor of $1/\pi$.}, are equal.}}

To get a $\pi^2$ sequence  we have to go a step further and use that in the limit $q\to\infty$  the overall binomial counting is unaffected by the trading of cosines: it follows that in (\ref{sonicebis}), or equivalently in (\ref{catiede}), setting $\rme^{ \rmi \pi A p/q}=1$, the infinite $A$ odd summation from minus to plus infinity of the    binomial multiple sums  \iffalse $\sum_{A=-\infty\atop  A\;{\it odd}}^{\infty}$\fi    necessarily converges  to
\begin{align}
\iffalse & \sum_{A=-\infty\atop  A\;{\it odd}}^{\infty}\nonumber\\&
 \sum_{k_3= -{rl_3/ 2}}^{{rl_3/2} 
 }\ldots \sum_{k_{j}= -{rl_j/2}}^{{rl_{j}/ 2}}
 {rl_1\choose {rl_1/ 2} +A/2+\sum_{i=3}^{j}(i-2)k_i}{rl_2\choose {rl_2/2} -A/2 - \sum_{i=3}^{j}(i-1)k_i}\prod_{i=3}^{j}{rl_i\choose {rl_i/ 2}+k_i}\nonumber\\&=\fi{r(l_1+l_2+\ldots+l_j)\choose r(l_1+l_2+\ldots+l_j)/2}\label{sonicesum}
\end{align}
  which is an integer  since $r$ is even\footnote{This can be directly checked, setting $\rme^{ \rmi \pi A p/q}=1$  in (\ref{sonicebis}), by first performing the $A$ summation using the generalized Chu-Vandermonde identity (\ref{chu}), then over the $k_i$'s by redefining  them appropriately (see \cite{nous}).}. 

 Using the  symmetry  $A\to-A$, consider now instead  the cumulative sum $ 2\sum_{A=1\atop  A\;{\it odd}}^{2m+1}$ of the same multiple binomial sums\iffalse(here written for the binomial multiple sums in (\ref{sonicebis}))\fi:
\iffalse
\begin{align}& 2\sum_{A=1\atop  A\;{\it odd}}^{2m+1}\nonumber\\&
 \sum_{k_3= -{rl_3/ 2}}^{{rl_3/2} 
 }\ldots \sum_{k_{j}= -{rl_j/2}}^{{rl_{j}/ 2}}
 {rl_1\choose {rl_1/ 2} +A/2+\sum_{i=3}^{j}(i-2)k_i}{rl_2\choose {rl_2/2} -A/2 - \sum_{i=3}^{j}(i-1)k_i}\prod_{i=3}^{j}{rl_i\choose {rl_i/ 2}+k_i}\nonumber
 \end{align}
 \fi
it is by construction of the form ${1/\pi^2}\;\;\times$ a rational  number  $a_{l_1, \ldots, l_j}(m)/ b_{l_1, \ldots, l_j}(m)$ with $a_{l_1, \ldots, l_j}(m)$ and $b_{l_1, \ldots, l_j}(m)$  becoming larger and larger  with $m$ going to infinity. It  then  follows  that
\be{ a_{l_1, \ldots, l_j}(m)\over b_{l_1, \ldots, l_j}(m)}\to_{m\to\infty} \pi^2{r(l_1+l_2+\ldots+l_j)\choose r(l_1+l_2+\ldots+l_j)/2}\nonumber\ee   
i.e., for any given set of $l_i$'s and $r$ even, we have constructed  a sequence of rational  numbers which converges when $m\to\infty$  to 
$\pi^2$ up to the overall binomial factor.

We can further sum  $ a_{l_1, \ldots, l_j}(m)/b_{l_1, \ldots, l_j}(m)$  over  all $g$-compositions  \cite{poly,brian} of  an integer $n$ (meaning that the sets  $l_1,l_2,\ldots,l_j$  are now viewed as the $g$-compositions of $n$, i.e., $l_1+l_2+\ldots+l_j=n$  with no more than $g-2$ zeroes in succession) with weight (see \cite{poly} for  its genesis) 
\bea
{c_g (l_1,l_2,\ldots,l_{j})} &=& {{{(l_1+\dots +l_{g-1}-1)!\over l_1! \cdots l_{g-1}!}~
\prod_{i=1}^{j-g+1} {l_i+\dots +l_{i+g-1}-1 \choose l_{i+g-1}}}}\nonumber\\
&=& {{{\prod_{i=1}^{j-g+1} (l_i + \dots + l_{i+g-1} -1)! \over \prod_{i=1}^{j-g} (l_{i+1} + \dots +l_{i+g-1} -1 )! } }\prod_{i=1}^j {1\over l_i!} }
\nonumber\eea
 to get the sequence of rational numbers
\be  {a_n(m)\over b_n(m)}= g n \sum_{l_1, l_2, \ldots, l_{j}\atop { \rm g-composition}\;{\rm of}\;n} c_g(l_1,l_2,\ldots,l_{j}) {a_{l_1, \ldots, l_j}(m)\over b_{l_1, \ldots, l_j}(m)}\nonumber\ee  
 By construction this sequence converges  when $m\to\infty$ to
\be{ a_{n}(m)\over b_{n}(m)}\to_{m\to\infty} \pi^2{r n\choose r n/2}{gn\choose n}\nonumber\ee

\section{Trading a single cosine \label{canne}}

In full  generality we could  trade in (\ref{simple0}) not  two cosines as in (\ref{sonicebis}), but any number of them for their
{\it shifted} binomial sums. Let us  here consider trading  the first cosine  only. In this case we cannot freely change the integration
range and each choice of range will yield a different results. We examine a couple of cases below.

\subsection{Integrating from $-1/2$ to $1/2$}
  \noindent Clearly 
\begin{align}
&\int_{-{1\over 2}}^{1\over 2} dt\sum_{k_1=-\infty\atop k_1\;{\it half}{\text -}{\it integer}}^{\infty}{rl_1\choose {rl_1/ 2}+k_1}\rme^{ 2\rmi \pi k_1 t}\prod_{i=2}^{j}\bigg(2\cos\big(\pi t -\pi(i-1)p/q\big)\bigg)^{rl_i}\label{soniceuno}
\end{align}
 is identical to (\ref{simple0})  by virtue of the {\it shifted} binomial theorem.
 We can again use the usual strategy to extract  from (\ref{soniceuno}) the relevant  binomial multiple sum:
binomial-expand each $i=2, \ldots, j$ cosine, see that the overall $\sum_{i=1}^j k_i$ multiplying $2\rmi\pi t$ in the exponential is now  half-integer because of $k_1$ being half-integer, define $A=-2\sum_{i=1}^j (i-1)k_i=-2\sum_{i=2}^j (i-1)k_i$, which is thus even, solve for $k_2=-A/2-\sum_{i=3}^{j}(i-1)k_i$ and finally integrate over $t$. We obtain
\begin{align}&\int_{-{1\over 2}}^{1\over 2} dt\prod_{i=1}^{j}\bigg(2\cos\big(\pi t -\pi(i-1)p/q\big)\bigg)^{rl_i}\nonumber\\=
&\sum_{A\;\text{even}}
 \, \rme^{ \rmi \pi A p/q}\sum_{k_1=-\infty\atop k_1\;{\it half}{\text -}{\it integer}}^{\infty}\sum_{k_3= -{rl_3/ 2} }^{{rl_3/2} 
 }\ldots \sum_{k_{j}= -{rl_j/2} }^{{rl_{j}/ 2}} \,{\sin\big(\pi(A/2+\sum_{i=1}^{j}(i-2)k_i)\big)\over \pi(A/2+\sum_{i=1}^{j}(i-2)k_i)} \nonumber\\&{rl_1\choose {rl_1/ 2}+k_1} {rl_2\choose {rl_2/ 2} -A/2-\sum_{i=3}^{j}(i-1)k_i}\prod_{i=3}^{j}{rl_i\choose {rl_i/ 2}+k_i}\label{cachauffe}
\end{align}
which is again $A\to -A$ symmetric (this a $\cos( \pi A p/q)$ expansion). Note that in (\ref{cachauffe}) the multiple binomial sums  weighted by $\rme^{ \rmi \pi A p/q}$  have the same form  as whose in (\ref{catiede}) with the caveat that in (\ref{cachauffe})  $A$ is even and $k_1$  half-integer whereas  in (\ref{catiede}) $A$  is odd and $k_1$  integer. (\ref{cachauffe})  is yet another rewriting of (\ref{sonice})  as again  a summation over  $A$ even of binomial  multiple sums   which are now $1/\pi^2\times$ a rational number  (there is an explicit  $1/\pi$   and another $1/\pi$   coming from the half-integer $k_1$ binomial).  For a given $A$ the multiple binomial sums   in (\ref{sonice}) and  (\ref{cachauffe}) match  one to one, meaning that an integer in (\ref{sonice}) is equal to  $1/\pi^2\;\;\times$ a rational number in  (\ref{cachauffe}) whose numerator and denominator become larger and larger  with the $k_1$ summation range    going to infinity.  Focusing instead on the cumulative sum $\sum_{k_1=-m+1/2}^{m+1/2}$ and multiplying it by $\pi^2$, it follows that its ratio  with the corresponding integer in (\ref{sonice})  yield  sequences of rational numbers converging  when $m\to\infty$ to $\pi^2$.

\subsection{Integrating from $0$ to $1$}

Let us now consider  instead of the integration range $-1/2$ to $1/2$ in (\ref{soniceuno}), the range  $0$ to $1$. The shifted  binomial theorem does hold anymore. The relevant binomial multiple sums can be extracted as usual by binomial-expanding each $i=2, \ldots, j$ cosine, defining $A=-2\sum_{i=1}^j(i-1)k_i$ which is  even, extracting $k_2=-A/2-\sum_{i=3}^{j}(i-1)k_i$  and finally  integrating over $t$
\begin{align}&\int_{0}^{1} dt\sum_{k_1=-\infty\atop k_1\;{\it half}{\text -}{\it integer}}^{\infty}{rl_1\choose {rl_1/ 2}+k_1}\rme^{ 2\rmi \pi k_1 t}\prod_{i=2}^{j}\bigg(2\cos\big(\pi t -\pi(i-1)p/q\big)\bigg)^{rl_i}\nonumber\\=
&\sum_{A\;\text{even}}
 \, \rme^{ \rmi \pi A p/q}\sum_{k_1=-\infty\atop k_1\;{\it half}{\text -}{\it integer}}^{\infty}\sum_{k_3= -{rl_3/ 2} }^{{rl_3/2} 
 }\ldots \sum_{k_{j}= -{rl_j/2} }^{{rl_{j}/ 2}} \, \rme^{-\rmi \pi(A/2+\sum_{i=1}^{j}(i-2)k_i)}{\sin\big(\pi(A/2+\sum_{i=1}^{j}(i-2)k_i)\big)\over \pi(A/2+\sum_{i=1}^{j}(i-2)k_i)}\nonumber\\&{rl_1\choose {rl_1/ 2}+k_1} {rl_2\choose {rl_2/ 2} -A/2-\sum_{i=3}^{j}(i-1)k_i}\prod_{i=3}^{j}{rl_i\choose {rl_i/ 2}+k_i}\nonumber
\end{align}
or, since $A$ being even and $k_1$  half-integer implies $\rme^ { 2\rmi\pi(A/2+\sum_{i=1}^{j}(i-2)k_i)}=-1$ so that  $\rme^ { -\rmi\pi(A/2+\sum_{i=1}^{j}(i-2)k_i)}\sin\big(\pi(A/2+\sum_{i=1}^{j}(i-2)k_i)\big)=-\rmi$
\begin{align}&\int_{0}^{1} dt\sum_{k_1=-\infty\atop k_1\;{\it half}{\text -}{\it integer}}^{\infty}{rl_1\choose {rl_1/ 2}+k_1}\rme^{ 2\rmi \pi k_1 t}\prod_{i=2}^{j}\bigg(2\cos\big(\pi t -\pi(i-1)p/q\big)\bigg)^{rl_i}\nonumber\\=
&-\rmi\sum_{A\;\text{even}}
 \, \rme^{ \rmi \pi A p/q}\sum_{k_1=-\infty\atop k_1\;{\it half}{\text -}{\it integer}}^{\infty}\sum_{k_3= -{rl_3/ 2} }^{{rl_3/2} 
 }\ldots \sum_{k_{j}= -{rl_j/2} }^{{rl_{j}/ 2}} \, {1\over \pi(A/2+\sum_{i=1}^{j}(i-2)k_i)}\nonumber\\&{rl_1\choose {rl_1/ 2}+k_1} {rl_2\choose {rl_2/ 2} -A/2-\sum_{i=3}^{j}(i-1)k_i}\prod_{i=3}^{j}{rl_i\choose {rl_i/ 2}+k_i}\label{cachauffebis}
\end{align}
which is now {\it anti-}symmetric under $A \to -A$,
since the denominator $\pi(A/2+\sum_{i=1}^{j}(i-2)k_i)$ changes sign when simultaneously exchanging $A$ for $-A$ and the $k_i$'s for $-k_i$. It follows that this $\rme^{ \rmi \pi A p/q}$  expansion  is in fact a $\sin( \pi A p/q)$ expansion. {In (\ref{cachauffebis}) the  binomial  multiple sums weighted by   $\rme^{ \rmi \pi A p/q}$  are again (up to a factor $\rmi$) of the form  $1/\pi^2\; \times$ a rational number (there is
an explicit $1/ \pi$ and another $1/\pi$ coming from the half-integer $k_1$ binomial) whose   numerator and denominator become larger and larger  with $k_1$ going to infinity.

{ Could we find  if any  a  sequence associated to these rational numbers converging to a power of $\pi$? To do so let us  perform the integral in (\ref{cachauffebis}) in yet another way by relying  again on the {\it shifted} binomial theorem. To do so, we split the integration range in  two intervals $[0,1/2]$
and $[1/2,1]$:

\noindent - In the first  interval the {\it shifted} binomial theorem holds,  so the integral  is  
\begin{align}\int_{0}^{1/2} dt\prod_{i=1}^{j}\bigg(2\cos\big(\pi t -\pi(i-1)p/q\big)\bigg)^{rl_i}\nonumber
 \end{align}
 
 \noindent - In the second  interval, we change variable $t\to t-1/2\in [0,1/2]$:
\begin{align}&\int_{1/2}^{1} dt\sum_{k_1=-\infty\atop k_1\;{\it half}{\text -}{\it integer}}^{\infty}{rl_1\choose {rl_1/ 2}+k_1}\rme^{ 2\rmi \pi k_1 t}\prod_{i=2}^{j}\bigg(2\cos\big(\pi t -\pi(i-1)p/q\big)\bigg)^{rl_i}\nonumber\\&=\int_{0}^{1/2} dt\sum_{k_1=-\infty\atop k_1\;{\it half}{\text -}{\it integer}}^{\infty}{rl_1\choose {rl_1/ 2}+k_1}\rme^{ 2\rmi \pi k_1 t}\rme^{ -\rmi \pi k_1 }\rme^{ 2\rmi \pi k_1 }\prod_{i=2}^{j}\bigg(2\cos\big(\pi t +\pi/2-\pi(i-1)p/q\big)\bigg)^{rl_i}\nonumber\\&=-\int_{0}^{1/2} dt\prod_{i=1}^{j}\bigg(2\sin\big(\pi t -\pi(i-1)p/q\big)\bigg)^{rl_i}\nonumber
 \end{align}
 the last line following from using the {\it shifted} binomial theorem and $k_1$ being  half-integer.

\noindent Altogether we obtain
\begin{align}&\int_{0}^{1} dt\sum_{k_1=-\infty\atop k_1\;{\it half}{\text -}{\it integer}}^{\infty}{rl_1\choose {rl_1/ 2}+k_1}\rme^{ 2\rmi \pi k_1 t}\prod_{i=2}^{j}\bigg(2\cos\big(\pi t -\pi(i-1)p/q\big)\bigg)^{rl_i}\nonumber\\&=\int_{0}^{1/2} dt\bigg(\prod_{i=1}^{j}\bigg(2\cos\big(\pi t -\pi(i-1)p/q\big)\bigg)^{rl_i}-\prod_{i=1}^{j}\bigg(2\sin\big(\pi t -\pi(i-1)p/q\big)\bigg)^{rl_i}\bigg) \nonumber
 \end{align}
from which the relevant binomial multiple sums can be extracted as above (binomial-expand each $i=1, \ldots, j$ cosine  and sine, ...) \iffalse usual by binomial-expanding each cosine and sine, defining $A=-2\sum_{i=1}^j(i-1)k_i$ which is  even, extracting $k_2=-A/2-\sum_{i=3}^{j}(i-1)k_i$  and finally integrating over $t$ \fi to obtain
\begin{align}&\int_{0}^{1} dt\sum_{k_1=-\infty\atop k_1\;{\it half}{\text -}{\it integer}}^{\infty}{rl_1\choose {rl_1/ 2}+k_1}\rme^{ 2\rmi \pi k_1 t}\prod_{i=2}^{j}\bigg(2\cos\big(\pi t -\pi(i-1)p/q\big)\bigg)^{rl_i}
\nonumber\\&=-\rmi\sum_{A\;\text{even}}
 \, \rme^{ \rmi \pi A p/q}\sum_{k_1=-rl_1/ 2}^{rl_1/ 2}\sum_{k_3= -{rl_3/ 2} }^{{rl_3/2} 
 }\ldots \sum_{k_{j}= -{rl_j/2} }^{{rl_{j}/ 2}} \, {1-\cos\bigg(\pi \big(A/2+\sum_{i=1}^{j}(i-2)k_i\big)\bigg)\over \pi\big(A/2+\sum_{i=1}^{j}(i-2)k_i\big)}\nonumber\\&{rl_1\choose {rl_1/ 2}+k_1} {rl_2\choose {rl_2/ 2} -A/2-\sum_{i=3}^{j}(i-1)k_i}\prod_{i=3}^{j}{rl_i\choose {rl_i/ 2}+k_i}
\label{0000}
 \end{align}
In (\ref{0000})   the multiple binomial sums  weighted by $ \rme^{ \rmi \pi A p/q}$  are now of the form $1/\pi \;\;\times$ a rational 
number\footnote{\noindent Since $A$ is even and all the $k_i$'s are integers,   $\pi\big(A/2+\sum_{i=1}^{j}(i-2)k_i\big)$  in the denominator can vanish  so that $1-\cos\bigg(\pi \big(A/2+\sum_{i=1}^{j}(i-2)k_i\big)\bigg)$ in the numerator also vanishes: the indeterminate ratio has to be  understood as also vanishing.}  (up to a factor $i$). Since  they  match one to one  those in (\ref{cachauffebis}),   one can  again construct  a sequence of rational numbers converging now to $\pi$, since the former   being  when multiplied by $\pi^2$  a rational number whose numerator and denominator become larger and larger with the $k_1$ summation range  going to infinity --so here one again focuses on the cumulative sum $\sum_{k_1=-m+1/2}^{m+1/2}$-- and the latter being when multiplied by $\pi$ a rational number, their  ratio is a rational number which necessarily converges when $m\to\infty$ to $\pi$.}

\section{Conclusions}
\noindent
We have explored the relation of trigonometric sums with shifted summation variables to corresponding trigonometric integrals. Clearly this is the tip
of a big iceberg that depends on how we wish to extend or deform these sums. E.g., we could consider yet other  tradings of  cosines for  shifted binomial sums, as for example with  {\bf four} {\it shifted} binomial sums (i.e., an even number of { shifted} sums, so integrating over $[0,1]$ is the same as integrating over $[-1/2,1/2]$)}
\begin{align}
&\int_{-1/2}^{1/2} dt\sum_{k_1=-\infty\atop k_1\;{\it half}{\text -}{\it integer}}^{\infty}\sum_{k_2=-\infty\atop k_2\;{\it half}{\text -}{\it integer}}^{\infty}\sum_{k_3=-\infty\atop k_3\;{\it half}{\text -}{\it integer}}^{\infty}\sum_{k_4=-\infty\atop k_4\;{\it half}{\text -}{\it integer}}^{\infty}{rl_1\choose {rl_1/ 2}+k_1}{rl_2\choose {rl_2/ 2}+k_2}{rl_3\choose {rl_3/ 2}+k_3}{rl_4\choose {rl_4/ 2}+k_4}\nonumber\\&\rme^{ 2\rmi \pi k_1 t}\rme^{ 2\rmi \pi k_2(t- p/q)}\rme^{ 2\rmi \pi k_3(t- 2p/q)}\rme^{ 2\rmi \pi k_4(t- 3p/q)}\prod_{i=5}^{j}\bigg(2\cos\big(\pi t -\pi(i-1)p/q\big)\bigg)^{rl_i}\nonumber
\end{align}
which would  again yield in the limit $q\to\infty$ the  overall binomial counting ${r(l_1+l_2+\ldots+l_j)\choose r(l_1+l_2+\ldots+l_j)/2}$. 
Proceeding as above we would obtain that this trigonometric integral rewrites as 
\begin{align}
 &\sum_{A=-\infty\atop  A\;{\it even}}^{\infty}
 \, \rme^{ \rmi \pi A p/q}\, \sum_{k_3=-\infty\atop k_3\;{\it half}{\text -}{\it integer}}^{\infty}\sum_{k_4=-\infty\atop k_4\;{\it half}{\text -}{\it integer}}^{\infty}\sum_{k_5= -{rl_5/ 2}}^{{rl_5/2} 
 }\ldots \sum_{k_{j}= -{rl_j/2}}^{{rl_{j}/ 2}}\nonumber \\
&\hskip 0.5cm {rl_1\choose {rl_1/ 2} +A/2+\sum_{i=3}^{j}(i-2)k_i}{rl_2\choose {rl_2/2} -A/2 - \sum_{i=3}^{j}(i-1)k_i}\prod_{i=3}^{j}{rl_i\choose {rl_i/ 2}+k_i} \label{000}
\end{align}  where $A$ is necessarily even. One could  take advantage of the shifted binomial theorem and rewrite the first shifted binomial sum as a cosine  to obtain yet another expression of the  $\rme^{ \rmi \pi A p/q}$ expansion (\ref{000}). We could also use the overall binomial sum rule. 
 These manipulations would lead to rational sequences for {$\pi^4$ --since the first four binomial entries in (\ref{000}) are  half-integers-- albeit here with two infinite summations over the half-integers $k_3$ and $k_4$}. 
 
It is  clear that this pattern generalizes to any number of shifted binomial sums. Extensions of sums to other rational values of $A$ would also be
possible, using the shifted binomial theorem, and would lead to similarly generalized results.

An interesting implication of our results would be in the possible connection of the full sums, over both $A$ and the $g$-compositions $l_i$, and interacting spin systems. The connection arises by mapping each $g$-composition of $n$ with a configuration of $n-1$ $g$-level systems (e.g.,
spin-$(g-1)/2$ $SU(2)$ spins or fundamental $SU(g)$ spins). The corresponding spin dynamics are implied by the sums, interpreted as spin partition functions. The generalized sums considered in this work would still correspond to spin systems, but with different, nontrivial couplings. 

Finally, the shifted binomial theorem itself could be explored and mined for applications, irrespective of any random walk connection. In particular, it could be used to express periodic functions as a sum of shifted Fourier frequencies, rather than the standard Fourier sum. This could be useful, e.g., for signal processing  where the shifted frequency expansion may converge faster, or be more revealing of the properties of the signal.

These and similar considerations are left for future work.
\vskip 0.4cm

%{\bf Acknolwledgments:}

{\section{Appendix}}

Let us start from the shifted binomial theorem (\ref{shiftybis})
\be
\left(2\cos({\pi t})\right)^l ={\sum _{k=-\infty\atop {\it integer}\;{\it or}\;{\it half}{\text -}{\it integer}}^{\infty}}{l \choose {l/ 2} + k + s} \, \rme^{2\rmi \pi (k+s) t} ~,~~ t  \in (-1/2 , 1/2 )
\nonumber\ee
{where $k$ is integer if $l$ is even and half-integer if $l$ is odd}  and focus on $s$ a rational number.

Again take $t=0$: we obtain
\be
2^l =\sum _{k=-\infty\atop {\it integer}\;{\it or}\;{\it half}{\text -}{\it integer}}^{\infty}{l \choose {l/ 2} + k + s}
\nonumber\ee
When $l$ is even i.e., $k$ integer, the shifted binomial ${l \choose {l/ 2} + k + s}$ is  a rational number up to the factor  $1/(\Gamma[s]\Gamma[1-s])=\sin(\pi s)/\pi$. It means that necessarily the sequence
\be2^{-l}\lim_{m\to\infty} \sum_{k=-m}^{m}{\pi\over\sin(\pi s)}{l\choose {l/ 2}+k+s}={\pi\over\sin(\pi s)}\nonumber \ee 
is rational and  converges to $ {\pi/\sin(\pi s)}$.

\noindent Likewise, when $l$ is odd, i.e., $k$ half-integer, \iffalse the shifted binomial ${l \choose {l/ 2} + k + s}$ is again  a rational number up to the factor   $\sin(\pi s)/\pi$. It means that necessarily \fi the sequence
\be 2^{-l}\lim_{m\to\infty} \sum_{k=-m-1/2}^{m-1/2}{\pi\over\sin(\pi s)}{l\choose {l/ 2}+k+s}={\pi\over\sin(\pi s)} \nonumber \ee 
is rational and  converges  to $ {\pi/\sin(\pi s)}$.

Upon integrating over t in the interval $[-1/2,1/2]$ one would get
\be
{l\choose l/2} =\sum _{k=-\infty\atop {\it integer}\;{\it or}\;{\it half}{\text -}{\it integer}}^{\infty}{l \choose {l/ 2} + k + s}{\sin\big(\pi (k+s)\big)\over \pi (k+s)}
\nonumber\ee
Altogether,
\begin{itemize}
\item when $l$ is even, the sequence
\be {(l/2)!^2\over l!}\lim_{m\to\infty}\sum_{k=-m}^{m}{\pi\over\sin(\pi s)} {l\choose {l/ 2}+k+s}{(-1)^k\over k+s}=\big({\pi\over\sin(\pi s)}\big)^2 \nonumber \ee 
is rational and  converges to $ \big({\pi/\sin(\pi s)}\big)^2 $.
\item when $l$ is odd, the sequence  \iffalse 
\be{(l/2)!^2\over l!}\lim_{m\to\infty}\sum_{k=-m-1/2}^{m-1/2}{ \Gamma(s)^2\Gamma(1-s)^2\cot(\pi s)\over \pi}{l\choose {l/ 2}+k+s}{\sin\big(\pi (k+s)\big)\over \pi (k+s)}={ \Gamma(s)^2\Gamma(1-s)^2\cot(\pi s)\over \pi} \nonumber \ee 
is rational and  converges to $ { \Gamma(s)^2\Gamma(1-s)^2\cot(\pi s)/ \pi}$. 
\fi
\be {(l/2)!^2\over l!}{1\over \pi}\lim_{m\to\infty}\sum_{k=-m-1/2}^{m-1/2}{\pi\over\sin(\pi s)} 
{l\choose {l/ 2}+k+s}{(-1)^{k-1/2}\over k+s}={\pi\over\sin(\pi s)\cos(\pi s)} \nonumber \ee 
($l$ being odd there is an additional $1/\pi$ due to  ${(l/2)!^2/ l!}$) is rational and converges to $ \pi/\big( \sin(\pi s)\cos(\pi s)\big) $.
\end{itemize}

}

\end{document}